\documentclass[12pt,a4paper,DIV12]{scrartcl}
\usepackage[T1]{fontenc}
\usepackage{lmodern}
\usepackage[british]{babel}
\usepackage{amsmath}
\usepackage{amssymb}
\usepackage{amsfonts}
\usepackage{color}
\usepackage{float}
\usepackage{hyperref}

\newcommand{\tr}{\ensuremath{\mathrm{tr}}}
\newcommand{\str}{\ensuremath{\mathrm{str}}}

\newcommand{\I}{\ensuremath{\mathrm{i}\hspace{1pt}}}
\newcommand{\arxiv}[1]{arXiv:\,\href{http://arxiv.org/abs/#1}{{\tt #1}}}
\newcommand{\aetap}{\text{a--}\eta'}
\newcommand{\api}{\text{a--}\pi}
\begin{document}
\title{
  {\vspace{-20mm}\normalsize
   \hfill\parbox[b][30mm][t]{35mm}{\textmd{MS-TP-14-10}}}\\[-18mm]
The mass of the adjoint pion in $\mathcal{N}=1$ supersymmetric Yang-Mills theory
\vspace*{3mm}}
\author{G.~M\"unster, H.~St\"uwe\\
\textit{\large Universit\"at M\"unster, Institut f\"ur Theoretische Physik}\\
\textit{\large Wilhelm-Klemm-Str.~9, D-48149 M\"unster, Germany}
\vspace*{5mm}}

\date{February 19, 2014}

\maketitle

\begin{abstract}
In Monte Carlo simulations of $\mathcal{N}=1$ supersymmetric Yang-Mills theory
the mass of the unphysical adjoint pion, which is easily obtained numerically,
is being used for the tuning to the limit of vanishing gluino mass. 
In this article we show how to define the adjoint pion in the framework
of partially quenched chiral perturbation theory and we derive a relation
between its mass and the mass of the gluino analogous to the Gell-Mann-Oakes-Renner
relation of QCD.
\end{abstract}
\vspace{5mm}

The $\mathcal{N}=1$ supersymmetric Yang-Mills theory (SYM) is the supersymmetric
extension of non-Abelian gauge theory. It describes gluons,
belonging to gauge group SU($N_c$), interacting with their superpartners,
the gluinos. The gluons are represented by non-Abelian gauge fields
$A_{\mu}(x) = A_{\mu}^{a}(x) T^{a}$, $a = 1, \dots, N_c^{2}-1$, 
where $T^{a}$ are the generators of the gauge group.
The gluinos
$\lambda(x) = \lambda^{a}(x) T^{a}$ are spin 1/2 Majorana fermions.
They are in the adjoint representation of the gauge group and their
gauge covariant derivative is given by
$\displaystyle
\mathcal{D}_{\mu} \lambda^{a} = \partial_{\mu} \lambda^{a} +
g\,f_{abc} A^{b}_{\mu} \lambda^{c}$.
The (on-shell) Lagrangian of SYM is
\begin{equation}
\mathcal{L}=\tr\left[-\frac{1}{2} F_{\mu\nu}F^{\mu\nu} 
+ \I \bar{\lambda} \gamma^\mu \mathcal{D}_{\mu} \lambda
- m_g \bar{\lambda}\lambda \right] \,,
\end{equation}
where $F_{\mu\nu}$ is the non-Abelian field strength.
A gluino mass term has been added, that breaks supersymmetry softly.
In the limit $m_g = 0$ the action is invariant under a supersymmetry transformation.

In recent years Monte Carlo simulations of SYM have been performed in order
to study its non-perturbative properties, in particular to determine
the spectrum of low-lying bound states; 
see \cite{Bergner:2013nwa,Bergner:2013jia} and references therein.
In these calculations the lattice regularisation of SYM proposed by Curci and
Veneziano \cite{Curci:1986sm} has been employed. Here the
gluinos are represented by Wilson fermions. Both supersymmetry and chiral
symmetry are broken by the lattice discretisation. 
In order to approach these symmetries in the continuum limit, a fine-tuning of
the bare gluino mass parameter is necessary
\cite{Curci:1986sm,Suzuki:2012pc}.
As in the Curci-Veneziano formulation the gluino mass term is not protected 
against additive renormalisation, the point of vanishing gluino mass is not given
a priori, but has to be determined on the basis of suitable observables.

One possibility to determine the gluino mass is to employ the lattice 
supersymmetric Ward identities as discussed in \cite{Farchioni:2001wx}.
Another, numerically much easier way is to monitor the mass of the
adjoint pion ($\api$),
which is the pion in the corresponding theory with two Majorana fermions
in the adjoint representation. 
The $\api$ is not a physical particle in SYM, which only contains one
Majorana fermion. Its mass can, however, obtained unambiguously from
the corresponding correlation function, which
is obtained as follows.
One of the mesonic bound states described by SYM is the so-called
adjoint $\eta'$ ($\aetap$), which is a colourless pseudoscalar particle
with interpolating field $\bar\lambda(x) \gamma_5 \lambda(x)$.
Its correlation function contains connected and disconnected fermionic contributions. The connected part
\begin{equation}
C(x,y)
= \langle \tr_{sc}[\gamma_5 (\gamma^\mu \mathcal{D}_{\mu})^{-1}(x,y)
\gamma_5 (\gamma^\mu \mathcal{D}_{\mu})^{-1}(y,x)] \rangle,
\end{equation}
where $\tr_{sc}$ denotes a trace over Dirac and colour indices,
yields the $\api$-correlation function.

The adjoint pion mass is expected to vanish in the limit of a massless gluino
according to 
\begin{equation}
\label{GOR}
m^2_{\api} \propto m_g, 
\end{equation}
analogous to the Gell-Mann-Oakes-Renner (GOR)
relation of QCD,
as has been argued on the basis of
the OZI-approximation of SYM \cite{Veneziano:1982ah}.
Indeed, numerical investigations of both
the gluino mass from supersymmetric Ward identities and
the adjoint pion mass \cite{Demmouche:2010sf} have shown that
the points of their vanishing are consistent with each other, and that
$m^2_{\api}$ is proportional to $m_g$.
In practice the $\api$ is being used for tuning since it
yields a more precise signal than the
supersymmetric Ward identities.

It is the purpose of this article to demonstrate that the adjoint
pion can be defined in a partially quenched setup,
in which the model is supplemented by a second species of gluinos and the
corresponding bosonic ghost gluinos, in the same way as for
one-flavour QCD \cite{Farchioni:2007dw}, and to show that the behaviour
indicated in Eq.~(\ref{GOR}) is indeed
found in partially quenched chiral perturbation theory.

Apart from the classical U(1)$_A$ axial symmetry, which is anomalous in
the quantum theory, SYM does not have a continuous chiral symmetry.
Therefore it also does not show spontaneous chiral symmetry breaking and
does not have (pseudo-) Goldstone bosons like pions, whose masses would vanish
in the chiral limit. The symmetry can, however, be enhanced artificially
by adding additional flavours of gluinos $\lambda_i(x)$, $i=2,\dots, N$. 
If these additional gluinos were
dynamical, the resulting theory would be different from SYM and would not be
supersymmetric.
On the other hand, if the additional gluinos are {\em quenched}, which means
that they are not taken into account in the fermionic functional integral,
the dynamical content of the model is identical to SYM and the correlation
functions of the original fields are unchanged. This situation can be called
{\em partially quenched}. It can be described theoretically by the
introduction of bosonic ghost fermions \cite{Morel}, in our case ghost gluinos.
The contribution of the ghost gluinos exactly cancels the contribution
of the additional gluinos, and only the contribution of the original single
gluino remains.

In the partially quenched setup adjoint pions can be formed out of
the gluinos $\lambda_1 \equiv \lambda$ and $\lambda_2$ by means of
$\bar{\lambda}_i \gamma_5 (\tau_{\alpha})_{ij} \lambda_j$, where
$\tau_{\alpha}$ are the Pauli matrices.

Let us begin by considering the introduction of $N-1$ additional gluino fields. 
Of central importance for chiral perturbation theory and its partially
quenched variant is the flavour symmetry group
and its spontaneous breakdown. The fermionic kinetic term
$\bar{\lambda}_i \gamma^\mu \mathcal{D}_{\mu} \lambda_i$
in the Lagrangian has the same form as the corresponding quark term
in QCD. Due to the Majorana condition $\lambda = C \bar{\lambda}^T$
the left and right handed parts of the gluino fields are not independent
of each other and consequently the chiral symmetry group is not equal to 
$\textrm{SU}(N)_L \otimes \textrm{SU}(N)_R$ but to some subgroup of it.
If the hermitian generators of 
SU($N$) flavour transformations are denoted $T_{\alpha}$, a short
calculation reveals that the generators of the subgroup of
$\textrm{SU}(N)_L \otimes \textrm{SU}(N)_R$
consistent with the Majorana condition are given by
\begin{alignat}{2}
\text{those} \quad &T_{\alpha}, &\quad \text{for which} \quad 
&T_{\alpha} = - T_{\alpha}^{*},\\
\text{and those} \quad &T_{\alpha} \gamma_5, &\quad \text{for which} \quad 
&T_{\alpha} = T_{\alpha}^{*}.
\end{alignat}
They generate a subgroup isomorphic to SU($N$), which is the chiral
symmetry group of $N$ gluinos. Another way to view this group is to
write the gluinos in terms of two-component Weyl fermions $\chi$,
\begin{equation}
\lambda=
\begin{pmatrix}
\chi\\
-\epsilon\chi^{*}
\end{pmatrix},
\end{equation}
where $\epsilon$ is the two-dimensional antisymmetric spinor-metric,
and to represent the kinetic term as
\begin{equation}
\mathcal{L}_g = \chi_{i}^{\dagger}\, \bar{\sigma}^{\mu} \mathcal{D}_{\mu} \chi_i \,.
\end{equation}
From this expression one
directly sees that SU($N$) transformations of the Weyl fields $\chi_i$
leave the kinetic term invariant.

The gluino mass term proportional to $\bar{\lambda}_i \lambda_i$ is invariant
under the subgroup $H$ of SU($N$), which is generated by the
$N (N-1)/2$ imaginary $T_{\alpha}$, i.e.\ $T_{\alpha} = - T_{\alpha}^{*}$.
The corresponding group elements $h = \exp (\I h_{\alpha} T_{\alpha})$
are real orthogonal matrices, and we see that $H = \textrm{SO}(N)$.

Assuming that the chiral symmetry group SU($N$) of gluinos is spontaneously
broken, accompanied by a non-vanishing gluino condensate 
$\langle \bar{\lambda}_i \lambda_j \rangle \propto \delta_{ij}$,
the breakdown from $G = \textrm{SU}(N)$ to $H = \textrm{SO}(N)$ is precisely
one of the three scenarios for spontaneous symmetry breakdown discussed by Peskin
\cite{Peskin80}, adapted to Majorana fermions \cite{Halasz:1995qb}.
The Goldstone boson manifold is the coset space $G/H$. Chiral perturbation
theory is based on an effective field theory for Goldstone bosons.
For its formulation a suitable parameterisation of the Goldstone boson manifold is needed.
The general procedure for formulating effective theories and finding the
associated effective Lagrangians has been developed in Ref.~\cite{CWZ69} and leads
to nonlinear representations of the chiral symmetry group.

As for the discussion of the adjoint pion in SYM it is sufficient to consider only
one additional gluino, we shall consider the case $N=2$ for definiteness 
in the following. So we have $G = \textrm{SU}(2)$ and 
$H = \textrm{SO}(2) = \textrm{U}(1)$. 
The subgroup $H$ is generated by $T_2 = \sigma_2 /2$.
The homogeneous space $\textrm{SU}(2)/\textrm{U}(1)$ is isomorphic to the sphere
$S^{\,2}$. Therefore it would be possible to represent the Goldstone boson field
by a real unit vector field $\vec{n}(x)$ and formulate the effective Lagrangian as 
a non-linear $\sigma$-model for $\vec{n}(x)$. In our case there is, however, 
another way, which is more convenient for explicit calculations.

Abstractly defined, the coset space $G/H$ is equal to the set of cosets
$gH$ with $g \in G$. Every element of SU(2) (apart from exceptional points)
can be parameterised uniquely as
\begin{equation}
g = \exp (\I\alpha_1 T_1 + \I\alpha_3 T_3)\, \exp (\I\alpha_2 T_2)
\doteq u h
\end{equation}
with real parameters $\alpha_k$.
Therefore the elements of the coset space $G/H$ can be parameterised as
\begin{equation}
u = \exp (\I\alpha_1 T_1 + \I\alpha_3 T_3) .
\end{equation}
These matrices are unitary and symmetric, $u^T = u$.
One could now set up chiral perturbation theory by introducing the field
$\alpha(x) = \alpha_1(x) T_1 + \alpha_3(x) T_3$
via
\begin{equation}
u(x) = \exp \left(\I \frac{\alpha(x)}{F}\right)
\end{equation}
with a dimensionful ``decay constant'' $F$. The transformation law of $u(x)$
under the chiral group SU(2) would, however, be complicated.
Instead we introduce the field
\begin{equation}
U(x) = \exp \left(\I \frac{\phi(x)}{F}\right)
\end{equation}
through
\begin{equation}
U(x) = u(x)^2 = u(x) u(x)^T .
\end{equation}
This matrix valued field transforms in a simple way under SU(2), namely
\begin{equation}
U(x) \rightarrow U'(x) = V U(x) V^T\,, \qquad V \in \textrm{SU}(2),
\end{equation}
similar to the case of QCD.
Examples for invariant expressions are
$\tr (A B^{\dagger})$ and $\tr (A B^{\dagger} C D^{\dagger})$ with
$A, B, C, D$ being derivatives of $U(x)$.

The effective Lagrangian can now be constructed in a standard way.
Let
\begin{equation}
\chi = 2 B_0 \mathcal{M} = 2 B_0 m_g \mathbf{1}
\end{equation}
be the mass term. The leading order effective Lagrangian is given by
\begin{equation}
\mathcal{L}_{2} 
= \frac{F^2}{4} \tr ( \partial_{\mu} U \partial^{\mu} U^{\dagger})
+ \frac{F^2}{4} \tr ( \chi U^{\dagger} + U \chi ^{\dagger})
\end{equation}
as for QCD. The next to leading order terms can be taken over from Gasser and
Leutwyler \cite{GL1}.

Let us now return to SYM, where in addition to the original sea gluino
the additional valence gluino is introduced in a
partially quenched manner. It is therefore accompanied by a ghost
gluino $\rho(x)$, having the same Lorentz transformation properties, but being a
boson. The chiral symmetry group is enhanced to the graded group SU(2|1)
and the Goldstone boson field $\phi(x)$, appearing
in the chiral field $U(x)$, is now a graded $3 \times 3$ matrix valued field,
\begin{equation}
\phi = 
\begin{pmatrix}
\phi_{ss} & \phi_{sv} & \phi_{sg} \\
\phi_{vs} & \phi_{vv} & \phi_{vg} \\
\phi_{gs} & \phi_{gv} & \phi_{gg}
\end{pmatrix}
\end{equation}
where the labels $s$, $v$ and $g$ stand for sea, valence and ghost.
For a graded matrix
\begin{equation}
M = 
\begin{pmatrix}
A & B \\
C & D
\end{pmatrix},
\end{equation}
where $A$ is a $2 \times 2$ matrix and $D$ a $1 \times 1$ matrix
(number), the supertrace is defined by
\begin{equation}
\str (M) = \tr (A) - \tr (D).
\end{equation}
The leading order effective Lagrangian reads
\begin{equation}
\mathcal{L}_{2}^{PQ} 
= \frac{F^2}{4} \str ( \partial_{\mu} U \partial^{\mu} U^{\dagger})
+ \frac{F^2}{4} \str ( \chi U^{\dagger} + U \chi ^{\dagger}) .
\end{equation}
Based on the effective Lagrangian the masses of the pseudo-Goldstone bosons
can be calculated in partially quenched chiral perturbation theory
\cite{Bernard:1993sv,Sharpe:1997by}.
We have calculated the masses in next-to-leading order.
Whereas the tree-level contributions are similar to the ones in QCD, the loop
contributions differ due to the different group structure.
The adjoint pion is represented by $\phi_{sv}$. For its mass $m_{\api} = M_{sv}$
we find
\begin{equation}
M_{sv}^2 = 2 B_0 m_g + \frac{(2 B_0 m_g)^2}{F^2} (30 L_8 - 2 L_4 - 7 L_5 + 8 L_6),
\end{equation}
with Gasser-Leutwyler low-energy coefficients $L_i$. Interestingly the loop 
contribution vanishes so
that no chiral logarithm appears in $M_{sv}^2$.
For small $m_g$ we recognise the desired GOR-relation
\begin{equation}
m^2_{\api} = 2 B_0 m_g .
\end{equation}

\section*{Acknowledgements}

We thank Yigal Shamir for helpful discussions.


\end{document}